# Reproducible evaluation of diffusion MRI features for automatic classification of patients with Alzheimer's disease

Junhao Wen[a,b,c,d,e], Jorge Samper-González,[e,a,b,c,d], Simona Bottani[e,a,b,c,d,], Alexandre Routier[e,a,,b,c,d,f], Ninon Burgos[a,b,c,d,e], Thomas Jacquemont[a,b,c,d,e], Sabrina Fontanella[a,b,c,d,e], Stanley Durrleman[e,a,b,c,d], Stéphane Epelbaum[a,b,c,d,e,g], Anne Bertrand[a,b,c,d,e,h*], Olivier Colliot[a,b,c,d,e,i] , for the Alzheimer's Disease Neuroimaging Initiative[1]

[a]Institut du Cerveau et de la Moelle épinière, ICM, F-75013, Paris, France

[b]Inserm, U 1127, F-75013, Paris, France

[c]CNRS, UMR 7225, F-75013, Paris, France

[d]Sorbonne Université, F-75013, Paris, France

[e]Inria, Aramis project-team, F-75013, Paris, France

[f]Institut du Cerveau et de la Moelle épinière, ICM, FrontLab, F-75013, Paris, France

[g]AP-HP, Hôpital de la Pitié Salpêtrière, Institute of Memory and Alzheimer's Disease (IM2A), Centre of excellence of neurodegenerative disease (CoEN), Department of Neurology, F-75013, Paris, France.

[h]AP-HP, Hôpital Saint-Antoine, Department of Radiology, F-75013, Paris, France

[i]AP-HP, Hôpital de la Pitié Salpêtrière, Departments of Neurology and Neuroradiology, F-75013, Paris, France

* Deceased, March 2nd, 2018

[1] Data used in preparation of this article were obtained from the Alzheimer's Disease Neuroimaging Initiative (ADNI) database (adni.loni.usc.edu). As such, the investigators within the ADNI contributed to the design and implementation of ADNI and/or provided data but did not participate in analysis or writing of this report. A complete listing of ADNI investigators can be found at: http://adni.loni.usc.edu/wp-content/ uploads/how_to_apply/ADNI_Acknowledgement_List.pdf

Correspondence to:

Olivier Colliot, PhD - olivier.colliot@upmc.fr

Junhao Wen - junhao.wen89@gmail.com

ICM – Brain and Spinal Cord Institute

ARAMIS team

Pitié-Salpêtrière Hospital

47-83, boulevard de l'Hôpital, 75651 Paris Cedex 13, France

ORCID number:

Olivier Colliot: 0000-0002-9836-654X

Junhao Wen: 0000-0003-2077-3070

# Abbreviations

AD: Alzheimer's disease
CN: normal control
MCI: Mild Cognitive Impairment
sMCI: stable MCI
pMCI: progressive MCI
CV: cross-validation
ROI: region of interest
MMSE: mini-mental state examination
CDR: clinical dementia rating
QC: quality check
OMH: optimal margin hyperplane
T1w MRI: T1-weighted MRI
dMRI: diffusion MRI
fMRI: functional MRI
FA: fractional anisotropy
MD: mean diffusivity
RD: radial diffusivity
AD: axial diffusivity
MO: mode of anisotropy
FS: feature selection
FR: feature rescaling
DTI: diffusion tensor imaging
MRI: magnetic resonance imaging
PET: positron emission tomography
GM: gray matter
WM: white matter
BIDS: Brain Imaging Data Structure
RF: random forests
LR: logistic regression
NN: Nearest neighbors
NB: naive Bayes
ACC: accuracy
BA: balanced accuracy
AUC: area under the curve
SVM: support vector machine
RVM: relevance vector machine
LDA: linear discriminant analysis
ADNI: Alzheimer's Disease Neuroimaging Initiative
EDSD: European DTI Study on Dementia
SMA: Sydney Memory and Aging
RRMC: Research and Resource Memory
HSA: Hospital de Santiago Apostol
PRODEM: Prospective Registry on Dementia study
IDC: Ilsan Dementia Cohort
MCXWH: Memory Clinical at Xuan Wu Hospital
TJH: Tong Ji Hospital
MICPNU: Memory Impairment Clinic of Pusan National University Hospital
UHG: University Hospital of Geneva
DZNE: German Center for Neurodegenerative Diseases Rostock database
Local: private database
DUBIAC: Duke-UNC Brain Imaging and Analysis Center
NACC: National Alzheimer's Coordinating Center
NorCog: Norwegian registry for persons being evaluated for cognitive symptoms in specialized health care

# Abstract

Diffusion MRI is the modality of choice to study alterations of white matter. In past years, various works have used diffusion MRI for automatic classification of Alzheimer's disease. However, classification performance obtained with different approaches is difficult to compare because of variations in components such as input data, participant selection, image preprocessing, feature extraction, feature rescaling (FR), feature selection (FS) and cross-validation (CV) procedures. Moreover, these studies are also difficult to reproduce because these different components are not readily available. In a previous work (Samper-González et al. 2018), we propose an open-source framework for the reproducible evaluation of AD classification from T1-weighted (T1w) MRI and PET data. In the present paper, we first extend this framework to diffusion MRI data. Specifically, we add: conversion of diffusion MRI ADNI data into the BIDS standard and pipelines for diffusion MRI preprocessing and feature extraction. We then apply the framework to compare different components. First, FS has a positive impact on classification results: highest balanced accuracy (BA) improved from 0.76 to 0.82 for task CN vs AD. Secondly, voxel-wise features generally gives better performance than regional features. Fractional anisotropy (FA) and mean diffusivity (MD) provided comparable results for voxel-wise features. Moreover, we observe that the poor performance obtained in tasks involving MCI were potentially caused by the small data samples, rather than by the data imbalance. Furthermore, no extensive classification difference exists for different degree of smoothing and registration methods. Besides, we demonstrate that using non-nested validation of FS leads to unreliable and over-optimistic results: 5% up to 40% relative increase in BA. Lastly, with proper FR and FS, the performance of diffusion MRI features is comparable to that of T1w MRI. All the code of the framework and the experiments are publicly available: general-purpose tools have been integrated into the Clinica software package

(www.clinica.run) and the paper-specific code is available at: https://github.com/aramis-lab/AD-ML.

# 1. Introduction

Alzheimer's disease (AD), the most prevalent form of dementia, is expected to affect 1 out of 85 people in the world by the year 2050 (Brookmeyer et al. 2007). Neuroimaging offers the possibility to study pathological brain changes associated with AD in vivo (Ewers et al. 2011). The most common neuroimaging modalities used to study AD are T1-weighted (T1w) magnetic resonance imaging (MRI) and positron emission tomography (PET) with various tracers (Frisoni et al. 2010; Vemuri and Jack 2010). These techniques allow the study of different types of alterations in the gray matter (GM). However, while AD is often considered primarily a gray matter disease, white matter (WM) is also extensively altered. There has thus been an increased interest in using diffusion MRI to study alterations in WM as the disease progresses (Fellgiebel et al. 2006; Kantarci et al. 2001; Müller et al. 2005, 2007).

Over the past decades, there has been a strong interest in developing machine learning methods to assist the diagnosis and prognosis of AD based on neuroimaging data (Falahati et al. 2014; Sven Haller et al. 2011; Rathore et al. 2017). In particular, a large number of studies using machine learning have looked at the potential of diffusion MRI for AD classification (Cui et al. 2012; Dyrba, Barkhof, et al. 2015; Lella et al. 2017; M. Li et al. 2014; Maggipinto et al. 2017; Y. Xie et al. 2015). Several of these studies make use of the same publicly available dataset: the Alzheimer's Disease Neuroimaging Initiative (ADNI) (adni.loni.usc.edu). However, classification performance is not directly comparable across these studies because of differences in participant selection, feature extraction and selection, and performance metrics. It is thus difficult to know which approach performs best and which components of the method have the greatest influence on classification performance. We recently proposed a framework for the reproducible evaluation of machine learning algorithms in AD classification and demonstrated its use on PET and T1w MRI data (Samper-González et al. 2018). The framework is composed of tools for management of public datasets and in particular their conversion into the Brain Imaging Data Structure (BIDS) format (K. J. Gorgolewski et al. 2016), standardized preprocessing pipelines, feature extraction tools and classification algorithms as well as procedures for evaluation. This framework was devoted to T1w MRI and PET data.

In the present work, we extend this framework to diffusion MRI data and apply it to assess the influence of different components on classification performances. We first perform a systematic review of the previous works devoted to automatic classification of AD using diffusion MRI data. We then present the extension of the framework to diffusion MRI data, including conversion of ADNI diffusion MRI data into BIDS, and diffusion MR image preprocessing pipelines. We finally apply the framework to study the influence of various components on classification performance: feature type (voxel or regional features), imaging modality (T1w or diffusion MRI), data imbalance, smoothing strength, registration methods (single contrast or multicontrast registration), feature rescaling (FR) and feature selection (FS) strategy.

All the code (both of the framework and of the experiments) is publicly available: the general-purpose tools have been incorporated into Clinica (Routier et al. 2018), an open-source software platform that we developed for brain image analysis, and the paper-specific code is available at: https://github.com/aramis-lab/AD-ML.

## 2. State of the art

AD is associated with altered integrity of WM, in particular the loss of cellular barriers that constrain free water motion (S. Xie et al. 2006). The fact that DTI was designed to study WM microstructure has led to the hypothesis that DTI-based features can be used for AD classification (Selnes et al. 2013). In recent years, a large body of research has been published for classification of AD using diffusion MRI. Here, we provide a systematic review of works on this topic.

We performed an online search of publications concerning classification of AD using diffusion MRI. We included only publications in English, only original research publications (excluding review papers) and only peer-reviewed papers (either in journals or in conference proceedings), thereby excluding abstracts and preprints. We first searched on PubMed with the following search criteria: i) keywords: "((((classification diffusion MRI alzheimer's disease[Title/Abstract]) OR classification DTI alzheimer's disease[Title/Abstract]) OR diagnosis DTI alzheimer's disease[Title/Abstract]) OR diagnosis diffusion MRI alzheimer's disease[Title/Abstract]", ii) publication date: before the 31st October 2018, and iii) study species: humans. We identified 616 studies based on these criteria. Among these studies, 105 review papers were excluded. Based on the abstract, we then selected only papers devoted to AD classification and using at least diffusion MRI. This resulted in 18 studies. Secondly, another query was performed on Scopus with the following criteria: i) keywords: "(TITLE-ABS-KEY(classification OR diagnosis) AND TITLE-ABS-KEY((diffusion AND mri) OR dti) AND TITLE-ABS-KEY((alzheimer's OR alzheimer) AND disease))", and ii) publication date before the search day (the 31st October 2018). This resulted in 425 studies. We then excluded 104 review papers. Moreover, limiting to only peer-reviewed journals or conference proceedings resulted in 298 studies. Based on the abstract, we selected only papers devoted to AD classification and using at least diffusion MRI, resulting in 27 studies. After merging the studies found by both PubMed and Scopus, we obtained 32 studies. To complete this search, we also did a search on Google Scholar with keywords: "classification diffusion MRI alzheimer's disease" or "classification DTI alzheimer's disease" or "diagnosis DTI alzheimer's disease" or "diagnosis diffusion MRI alzheimer's disease". Two additional studies were included, resulting in a total of 34 studies which are presented in the current state-of-the-art section.

These 34 studies can be categorized according to the following criteria. i) *Studied modality.* While the majority used only diffusion MRI, some used multimodal data (combining diffusion MRI with T1w MRI or functional MRI for instance). ii) *Type of features.* We subdivided between papers using DTI metric features, such as fractional anisotropy (FA) and mean diffusivity (MD), and those using more advanced features, such as tract-based or network-based features. iii) *Classifiers.* The most commonly used are support vector machines (SVM) but random forests (RF), logistic regression (LR), nearest neighbours (NN) or naive Bayes (NB) were also used in some studies. iv) *Dataset.* The most commonly used dataset is the ADNI although it does not constitute an overwhelming majority, unlike for T1w-MRI or PET studies. This is probably because diffusion MRI was not present in ADNI1. v) *Classification task.* Some studies focused on the discrimination between AD patients and CN (cognitively normal) subjects while other tackled classification of patients with MCI (mild cognitive impairment) or prediction of progression to AD among MCI patients. vi) *FS approach.* A large proportion of the surveyed studies, denoted as *Non-nested* or *Unclear* in Table 1 and 2, may potentially suffer from FS bias, also known as non-nested FS strategy. It arises when FS is performed on the entire dataset and not within the cross-validation (CV) procedure, thus introducing data leakage. On the contrary, a nested FS is a procedure blind to the test data and embedded into the nested CV (Maggipinto et al. 2017). A summary of these characteristics for the different studies is presented in Tables 1 (for those using DTI metric features) and Table 2 (for connectivity or tractography features). Besides, if multimodal imaging or different type of features (i.e., DTI metric and more advanced features) were used in a study, we reported the results of the best performance.

**Table 1**. Summary of the studies using DTI metric features for AD classification.
a non-amnestic Mild Cognitive Impairment; b amnestic Mild Cognitive Impairment; c MCI-Aβ 42−; d MCI-Aβ 42+; e sd-aMCI, single domain amnestic MCI; f sd-fMCI, single domain frontal MCI; g md-aMCI, multiple domains amnestic MCI; h late MCI; i early MCI;
ACC: accuracy; BA: balanced accuracy; AUC: area under the curve;
--: not applicable. See the section of abbreviation for more details.

| Study | Subjects | | | Modality | Feature type | Classifier | Database | Performance | | | | Feature selection |
|---|---|---|---|---|---|---|---|---|---|---|---|---|
| | AD | MCI | CN | | | | | CN vs AD | CN vs MCI | sMCI vs pMCI | AD vs MCI | |
| (Ahmed et al. 2017) | 45 | 58 | 52 | dMRI, T1w | Hippocampal voxel MD | SVM | ADNI | BA=0.90 | BA=0.79 | -- | BA=0.73 | -- |
| (Schouten et al. 2016) | 77 | -- | 173 | dMRI, T1w, fMRI | Regional FA, MD | LR | PRODEM | AUC=0.95 | -- | -- | -- | -- |
| (Cui et al. 2012) | -- | 79b | 204 | dMRI, T1w | Regional FA | SVM | SMA | -- | ACC=0.71 | -- | -- | Nested |
| (Dyrba et al. 2013) | 137 | -- | 143 | dMRI | Voxel FA, MD | SVM | EDSD | ACC=0.83 | -- | -- | -- | Nested |
| (Lella et al. 2017) | 40 | -- | 40 | dMRI | Voxel FA, MD | SVM, RF, NB | ADNI | ACC=0.78 | -- | -- | -- | Nested |
| (M. Li et al. 2014) | 21 | -- | 15 | dMRI, T1w | Regional FA | SVM | TJH | ACC=0.94 | -- | -- | -- | Nested |
| (Y. Xie et al. 2015) | -- | 64b | 64 | dMRI, T1w | Voxel FA, MD | SVM | MCXWH | -- | ACC=0.84 | -- | -- | Nested |
| (Zhang and Liu 2018) | 48 | 39h, 75i | 51 | dMRI | Regional FA, MD, RD, AD | SVM, LR | ADNI | ACC=0.90 | -- | -- | -- | Nested |
| (Maggipinto et al. 2017) | 90 | 90 | 89 | dMRI | Voxel FA, MD | RF | ADNI | ACC=0.76 | ACC=0.60 | -- | -- | Nested |
| (Dyrba, Barkhof, et al. 2015) | -- | 35c, 42d | 25 | dMRI, T1w | Voxel FA, MD, MO | SVM | EDSD | -- | ACC=0.77d | ACC=0.68 | -- | Nested |
| (Dyrba, Grothe, et al. 2015) | 28 | -- | 25 | dMRI, T1w, fMRI | Regional FA, MD, MO | SVM | DZNE | AUC=0.89 | -- | -- | -- | Nested |
| (Jung et al. 2015) | 27 | 18 | -- | dMRI, T1w | Regional FA, MD | SVM | MICPNU | -- | -- | -- | ACC=0.87 | Non-nested |
| (Mesrob et al. 2012) | 15 | -- | 16 | dMRI, T1w | Voxel and regional FA, MD | SVM | RRMC | ACC=1 | -- | -- | -- | Non-nested |
| (O'Dwyer et al. 2012) | -- | 19a,14b | 40 | dMRI | Voxel FA, MD, RD, AD | SVM | EDSD | -- | ACC=0.93 | -- | -- | Non-nested |

| Study | | | | | | | | | | | |
|---|---|---|---|---|---|---|---|---|---|---|---|
| (S Haller et al. 2013) | -- | 18[e], 13[f], 35[g] | -- | dMRI | Voxel FA | SVM | Local | -- | -- | ACC=0.99[e,f] | -- | Non-nested |
| (Demirhan et al. 2015) | 43 | -- | 70 | dMRI | Voxel and regional FA | SVM | ADNI | ACC=0.88 | ACC=0.78 | -- | ACC=0.86 | Unclear |
| (Friese et al. 2010) | 21 | -- | 20 | dMRI, T1w | Voxel FA, MD | LR | Local | AUC=0.88 | -- | -- | -- | Unclear |
| (Graña et al. 2011) | 20 | -- | 25 | dMRI | Voxel FA, MD | SVM | HSA | ACC=1 | -- | -- | -- | Unclear |
| (Gao et al. 2015) | -- | 41 | 63 | dMRI, T1w, fMRI | Regional FA | -- | UHG | -- | ACC=0.85 | -- | -- | Unclear |
| (Lee et al. 2015) | 35 | 73 | 33 | dMRI | Voxel FA, MO | SVM | ADNI | ACC=0.88 | -- | -- | ACC=0.90 | Unclear |
| (Termenon et al. 2011) | 15 | -- | 20 | dMRI | Voxel FA, MD | SVM, RVM, NN | HSA | ACC=0.99 | -- | -- | -- | Unclear |

**Table 2.** Summary of the studies using tract-based or network-based features for AD classification.
a subjective decline MCI; b late MCI; c early MCI;
ACC: accuracy; BA: balanced accuracy; AUC: area under the curve;
--: not applicable. See the section of abbreviation for more information.

| Study | Subjects | | | Modality | Feature type | Classifier | Database | Performances | | | | Feature selection |
|---|---|---|---|---|---|---|---|---|---|---|---|---|
| | AD | MCI | CN | | | | | CN vs AD | CN vs MCI | sMCI vs pMCI | AD vs MCI | |
| (Amoroso et al. 2017) | 47 | -- | 52 | dMRI | Network measures | -- | ADNI | AUC=0.95 | -- | -- | -- | -- |
| (Doan et al. 2017) | 79 | 55, 30a | -- | dMRI | Tract measures, regional FA, MD, RD, AD | LR | NorCog | -- | -- | -- | AUC=0.71 | -- |
| (Cai et al. 2018) | 165 | -- | 165 | dMRI, T1w | Network measures | LDA | ADNI | ACC=0.85 | -- | -- | -- | Nested |
| (Ebadi et al. 2017) | 15 | 15 | 15 | dMRI | Network measures | LR, RF, NB, SVM, NN | -- | ACC=0.80 | ACC=0.70 | -- | ACC=0.80 | Nested |
| (Lella et al. 2018) | 40 | 30 | 52 | dMRI | Network measures | SVM | ADNI | AUC=0.77 | -- | -- | -- | Nested |
| (Prasad et al. 2015) | 38 | 38b,74c | 50 | dMRI | Network measures | SVM | ADNI | ACC=0.78 | -- | ACC=0.63 | -- | Nested |
| (Wee et al. 2012) | -- | 10 | 17 | dMRI, fMRI | Network measures | SVM | DUBIAC | -- | ACC=0.96 | -- | -- | Nested |
| (Q. Wang et al. 2018) | -- | 169 | 379 | dMRI, T1w | Network measures | SVM, RF | ADNI, NACC | -- | AUC=0.75 | -- | -- | Nested |
| (Zhu et al. 2014) | -- | 22 | 22 | dMRI, fMRI | Network measures | SVM | NACC | -- | ACC=0.95 | -- | -- | Non-nested |
| (Zhan et al. 2015) | 39 | 112 | 51 | dMRI | Network measures | LR | ADNI | BA=0.71 | BA=0.57 | -- | BA=0.69 | Non-nested |
| (Schouten et al. 2017) | 77 | -- | 173 | dMRI | Network measures, voxel FA, MD, RD, AD | LR | PRODEM | ACC=0.92 | -- | -- | -- | Non-nested |
| (Nir et al. 2015) | 37 | 113 | 50 | dMRI | Tract measures, FA, MD | SVM | ADNI | ACC=0.85 | ACC=0.79 | -- | -- | Non-nested |
| (Lee et al. 2013) | -- | 39 | 39 | dMRI | Tract measures, voxel and regional FA | SVM | ADNI | -- | ACC=1 | -- | -- | Unclear |

Twenty-one studies used DTI metrics as features (see details in Table 1). Among the DTI derived metrics, FA and MD were most frequently used (Dyrba, Barkhof, et al. 2015; Dyrba et al. 2013; Dyrba, Grothe, et al. 2015; Friese et al. 2010; Jung et al. 2015; Lella et al. 2017; Maggipinto et al. 2017; Mesrob et al. 2012; O'Dwyer et al. 2012; Schouten et al. 2016; Termenon et al. 2011; Y. Xie et al. 2015; Zhang and Liu 2018). Besides, radial diffusivity (RD), axial diffusivity (AD) and mode of anisotropy (MO) were also examined in some papers (Dyrba, Barkhof, et al. 2015; Dyrba, Grothe, et al. 2015; Lee et al. 2015; O'Dwyer et al. 2012; Zhang and Liu 2018). Voxel- and region-wise features were both used. For voxel-wise classification, all voxels from the segmented GM or WM were used. For region-wise classification, the mean value of DTI metric maps within regions of interest (ROI) were extracted using an anatomical atlas. The most commonly used atlases were the John Hopkins University (JHU) atlases (Hua et al. 2008). Ten studies adopted only diffusion MRI for AD classification (Demirhan et al. 2015; Dyrba et al. 2013; Graña et al. 2011; S Haller et al. 2013; Lee et al. 2015; Lella et al. 2017; Maggipinto et al. 2017; O'Dwyer et al. 2012; Termenon et al. 2011; Zhang and Liu 2018). The other eleven studies looked at the potential of multimodal MRI, for instance T1w MRI and diffusion MRI, for AD diagnosis and compared the performance cross modalities. For the DTI metric-based studies, SVM was most frequently used (Ahmed et al. 2017; Cui et al. 2012; Demirhan et al. 2015; Dyrba, Barkhof, et al. 2015; Dyrba et al. 2013; Dyrba, Grothe, et al. 2015; Graña et al. 2011; S Haller et al. 2013; Jung et al. 2015; Lee et al. 2015; Lella et al. 2017; M. Li et al. 2014; Mesrob et al. 2012; O'Dwyer et al. 2012; Termenon et al. 2011; Y. Xie et al. 2015; Zhang and Liu 2018).

Thirteen works demonstrated the usage of more complex features, such as tract-based or network-based features (see details in Table 2). In such approaches, tractography is used to extract white matter tracts from diffusion MRI data. To be reliable, such a procedure requires having high angular resolution diffusion imaging data. Then, tract-based approaches compute indices that characterize the tract, including tract volume, average FA/MD across the tract or more advanced features (Doan et al. 2017; Lee et al. 2013; Nir et al. 2015). Such indices are used as input of the classifier. In network-based features, the result of the tractography (also called the tractogram) is used to build a graph of anatomical connections. Usually, the gray matter is parcellated into a set of anatomical regions and the connectivity between two given regions is computed based on the tractogram. To that purpose, different measures have been used, including the number of fibers or the average FA along the connection. This results in a connectivity network which can be described through network-based measures. Such features characterize the local and global topology of the network and are fed to a classifier. Ten studies used network-based features derived from diffusion MRI for AD classification (Amoroso et al. 2017; Cai et al. 2018; Ebadi et al. 2017; Lella et al. 2018; Prasad et al. 2015; Schouten et al. 2017; Y. Wang et al. 2018; Wee et al. 2012; Zhan et al. 2015; Zhu et al. 2014).

There is a high variability in terms of classification performance across studies. The most frequently reported performance metric is the accuracy. Other metrics, such as balanced accuracy, area under the curve (AUC), specificity, and sensitivity, are also used. This makes it difficult to quantitatively compare classification performance across different studies. For DTI metric features, classification performance numerically ranges from 0.71 to 1 for task CN vs AD. With regard to the results across types of features, no consistency existed across studies. For instance, Nir et al observed that, in their study, the performance of MD outperformed FA (Nir et al. 2015). However, O'Dwyer et al reported higher accuracy for FA than MD in their experiments (O'Dwyer et al. 2012) and another study obtained comparable results for both metrics (Dyrba et al. 2013). Conflicting results were also reported for the comparison of different modalities. Mesrob et al obtained higher performance with T1w MRI than with diffusion MRI (Mesrob et al. 2012) while Dyrba et al came to the opposite conclusion (Dyrba, Barkhof, et al. 2015). For network- or tract-based features, the classification results numerically range from 0.71 to 0.95 for task CN vs AD, a range which is comparable to that obtained with DTI metrics. Of note, the numbers representing the general performance ranges are not for comparison purpose across studies.

In this work, we choose to focus on DTI metrics because: i) they are more simple than connectivity or tractography features; ii) they can be easily computed and can make use of standard diffusion MRI sequences, thus are more adapted to translation to clinical practice, iii) to date, there is no clear evidence that connectivity/tractography features lead to higher accuracies for AD classification; iv) conflicting results exist regarding the respective performance of different DTI metrics in this context.

## 3. Materials

The data used in the preparation of this article were obtained from the Alzheimer's Disease Neuroimaging Initiative database (ADNI) (adni.loni.usc.edu) (more details are presented in supplementary eMethod 1).

Five diagnosis groups were considered:

- CN: subjects who were diagnosed as CN at baseline;
- AD: subjects who were diagnosed as AD at baseline;
- MCI: subjects who were diagnosed as MCI, EMCI or LMCI at baseline;
- pMCI: subjects who were diagnosed as MCI, EMCI or LMCI at baseline, were followed during at least 36 months and progressed to AD between their first visit and the visit at 36 months;
- sMCI: subjects who were diagnosed as MCI, EMCI or LMCI at baseline, were followed during at least 36 months and did not progress to AD between their first visit and the visit at 36 months.

Naturally, all participants in the pMCI and sMCI groups are also in the MCI group. Note that the reverse is false, as some MCI subjects did not convert to AD but were not followed long enough to state whether they were sMCI or pMCI.

The diffusion-weighted images (DWIs) of ADNI were downloaded in October 2016. They all come from ADNI GO and ADNI2 phases. Two different acquisition protocols are described for DWIs: Axial DTI (images with "Sequence" field starting by "AX_DTI" and "Axial_DTI" in the file of "IDA_MR_Metadata_Listing.csv") and Enhanced Axial DTI (images with "Sequence" field equal to "Enhanced_Axial_DTI" in the file of "IDA_MR_Metadata_Listing.csv"). In total, Axial DTI were available for 1019 visits and Enhanced Axial DTI for 102 visits. Only Axial DTI images were available for the baseline visit (222). These DWIs were acquired with the following parameters: 35 cm field of view, 128×128 acquired matrix, reconstructed to a 256×256 matrix; voxel size: 1.35×1.35×2.7mm ; scan time = 9 min; 41 diffusion-weighted directions at b-value = 1000 s/mm$_2$ and 5 T2-weighted images (b-value = 0 s/mm$_2$, referred to as b0 image). Besides, each participant underwent a T1w MRI sequence with the following parameters: 256×256 matrix; voxel size = 1.2×1.0×1.0 mm ; TI = 400 ms; TR = 6.98 ms; TE = 2.85 ms; flip angle = 11°. We included the participants whose diffusion and T1w MRI scans were both available at baseline. We used quality check (QC) information provided by ADNI to select participants (see below Section 4.1). After converting images from DICOM to NIfTI format, four participants were excluded because of the lower image resolution (4.5×4.5×4.5mm). Moreover, QC was conducted on the results of the preprocessing pipeline (see below Section 4.2). Finally, 46 CN, 97 MCI, 54 sMCI, 24 pMCI and 46 AD were included.

Table 3 summarizes the demographics, and the MMSE and global CDR scores of the participants in this study.

**Table 3.** Summary of participant demographics, mini-mental state examination (MMSE) and global clinical dementia rating (CDR) scores. Values are presented as mean ± SD [range]. M: male, F: female

|  | N | Age | Gender | MMSE | CDR |
|---|---|---|---|---|---|
| CN | 46 | 72.7 ± 6.0 [59.8, 89.0] | 21 M / 25 F | 28.9 ± 1.4 [24,30] | 0: 46 |
| MCI | 97 | 72.9 ± 7.3 [55.0, 87.8] | 62 M / 35 F | 27.7 ± 1.7 [24,30] | 0.5: 97 |
| sMCI | 54 | 72.6 ± 7.7 [55.0, 87.8] | 21 M / 25 F | 28.0 ± 1.7 [24,30] | 0.5: 54 |
| pMCI | 24 | 74.2 ± 6.1 [56.5, 85.3] | 16 M / 8 F | 26.8 ± 1.4 [24,30] | 0.5: 24 |
| AD | 46 | 74.4 ± 8.4 [55.6, 90.3] | 28 M / 18 F | 23.4 ± 1.9 [20,26] | 0.5: 17; 1: 29; |

# 4. Methods

The classification framework is illustrated in Figure 1. It includes: tools for data management, image processing, feature extraction and selection, classification, and evaluation. Conversion tools allow an easy update of ADNI as new subjects become available. To facilitate future development and testing, the different components were designed in a modular-based architecture: processing pipelines using Nipype (K. Gorgolewski et al. 2011), and classification and evaluation tools using the scikit-learn[2] library (Pedregosa et al. 2011). Thus the objective measurement of the impact of each component on the results could be clarified. A simple command line interface is provided and the code can also be used as a Python library.

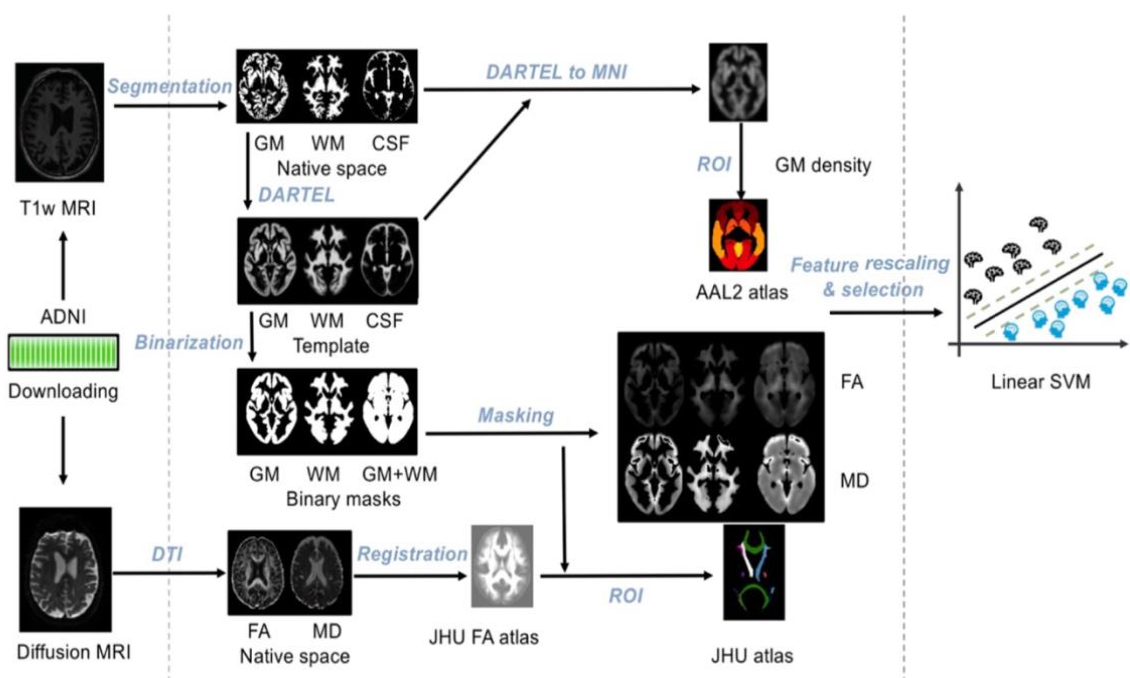

**Figure 1**. Overview of the framework.

## 4.1 Converting datasets to a standardized data structure

Public datasets, such as ADNI, are extremely useful to the research community. However, using the ADNI can be difficult because the downloaded raw data does not possess a clear and uniform organization. We thus proposed to convert ADNI data into the BIDS format (K. J. Gorgolewski et al. 2016), a community standard which allows storing multiple neuroimaging modalities as well as clinical and sociodemographic data. BIDS is based on a file hierarchy rather than on a database management system. It can thus be easily deployed in any research laboratory.

---

[2] http://scikit-learn.org

The ADNI to BIDS converter was originally developed in a previous study (Samper-González et al. 2018). Here, we extended that to diffusion MRI. The converter allows the user to automatically convert the raw dataset downloaded from the ADNI website to BIDS. The user has to download all the ADNI study data (tabular data in csv format) and the imaging data of interest. Importantly, the downloaded data must be kept exactly as they were downloaded. All conversion steps are then performed by the automatic converter, requiring no user intervention.

Details regarding conversion of clinical, sociodemographic and T1w MRI data can be found in (Samper-González et al. 2018). For DWIs, first, we selected from the file "IDA_MR_Metadata_Listing.csv", all entries containing "DTI" in the "Sequence" field. Images with a sequence name containing "Enhanced" were discarded. Then, "IMAGEUID" field was matched to corresponding "loni_image" field of 'MAYOADIRL_MRI_IMAGEQC_12_08_15.csv' file, to find QC information for each image. In cases where there existed several scans for a visit, we kept the one marked as selected (1 in 'series_selected' field of QC csv file). If there was no image marked as selected, then we chose the image with the best quality, (as specified in "series_quality" field, ranging from 1 to 4, 1 being excellent and 4 being unusable), excluding the images that failed QC (series_quality = 4). If there were several images for the same visit and QC information was not present, we chose the scan that was acquired first. No images were discarded at this step. Once paths for each of the selected images were gathered, the images in DICOM format were converted to NIfTI format using the *dcm2niix*[3] tool, or in case of error the *dcm2nii*[4] tool (X. Li et al. 2016). During the conversion from DICOM to NIfTI format, there were 4 images failing the conversion using both tools because of the lack of information of image header or the wrong format of b-value or b-vec file. Thus, they were manually discarded. Finally, the converted images in NIfTI format were organised by following the BIDS standard. Note that all these steps are automatically performed by the converter.

We also provide tools for subject selection according to the duration of follow up and the diagnose. In the present study, all the participants whose T1w MRI and diffusion MRI scans were available at baseline were included. Finally, we organized all the outputs of the experiments into a BIDS-inspired standardized structure.

## 4.2 Preprocessing pipelines

### 4.2.1 Preprocessing of T1w MRI

The image processing pipeline for T1w MRI was previously described in (Samper-González et al. 2018). In brief, the Unified Segmentation procedure(Ashburner and Friston 2005) is first used to simultaneously perform tissue segmentation (i.e., WM, GM and CSF tissue maps), bias correction and spatial normalization of the input image. Next, a group template is created using DARTEL (Ashburner 2007), from the subjects' tissue probability maps in native space obtained at the previous step. At this step, not only the group template, but also the deformation fields from each subject's native space into the DARTEL template space is obtained. Lastly, the DARTEL to MNI method (Ashburner 2007) is applied, providing a registration of the native space images into the MNI space: the transformation flow, for each subject, is a two-step combination from the native space into DARTEL template and finally into MNI space.

---

[3] https://github.com/rordenlab/dcm2niix
[4] https://www.nitrc.org/plugins/mwiki/index.php/dcm2nii:MainPage

The resulting transformation is applied to the subject's different tissue maps. Moreover, the DARTEL template generated its corresponding GM and WM tissue maps. These maps were then binarized with a probability threshold of 0.3 to obtain the tissue masks (i.e., GM, WM or GM+WM mask), which will be used in the following diffusion MRI pipeline.

### 4.2.2 Preprocessing of diffusion MRI

For each subject, all b0 images were rigidly registered to the first b0 image and then averaged as the b0 reference. The diffusion-related artifacts were corrected by the following steps. First, the raw DWIs were corrected for eddy current-induced distortions and subject movements by simultaneously modelling the effects of diffusion eddy currents and movements on the image using *eddy* tool (Andersson and Sotiropoulos 2016) from FMRIB Software Library (FSL) software (Jenkinson et al. 2012). Secondly, to correct for susceptibility-induced distortions, as fieldmap data were not available in ADNI1, ADNI GO or ADNI2, the T1w MRI was used instead in this context. The reference b0 image was firstly skull-stripped. To obtain the transformation flow from the native diffusion space into the T1w MRI space, the skull-stripped b0 image was registered to the T1w MRI with two sequential steps: first a rigid registration using FSL *flirt* tool and then a non-linear registration using SyN registration algorithm from ANTs (Avants et al. 2008). SyN is an inverse-consistent registration algorithm allowing EPI induced susceptibility artifacts correction (Leow et al. 2007). The resulting transformation was applied to the DWIs to correct for the susceptibility-induced distortions. Lastly, the DWI volumes were corrected for nonuniform intensity using the ANTs N4 bias correction algorithm (Tustison and Avants 2013). It estimates a single multiplicative bias field from the b0 reference image. Moreover, the diffusion weighting directions were appropriately updated (Leemans and Jones 2009). The implementation of these different steps is available in the *dwi-preprocessing-using-t1* pipeline of Clinica.

We performed QC on the results of the preprocessing pipeline. Specifically, we inspected the results for the presence of head motion artifacts and eddy current artifacts. Registration quality was also visually checked by overlapping the source image onto the target image. All preprocessed data were considered of acceptable quality.

The preprocessed DWIs were then fitted by DTI model to generate FA and MD maps using MRtrix (Tournier et al. 2012). In order to register individual DTI metric maps (i.e., FA and MD maps) into MNI space, two registration approaches were tested with the ANTs SyN algorithm (Avants et al. 2008): i) single contrast registration, where only the individual FA map was nonlinearly registered onto the JHU FA template; ii) multicontrast registration, where both the FA and b0 images were used for that purpose. The estimated nonlinear deformation with both approaches was finally applied to the MD maps to have all FA and MD maps in the same space. This is implemented in the *dwi-dti* pipeline of Clinica.

As the registration quality is crucial for the following analysis, we performed another QC step: the registered DTI metric maps were visually checked by overlapping the source image onto the target image. All images were considered of acceptable quality.

## 4.3 Feature extraction

We extracted two types of features: voxel-wise and regional features. After image processing, all T1w MRI and diffusion MRI are in MNI space and we have a voxel-wise correspondence across subjects. Voxel-wise features simply correspond to all the voxels in GM for T1w MRI. In order to extract the DTI-based voxel-wise features, FA and MD maps were masked using the tissue masks (i.e., WM, GM and GM+WM tissue binarized masks) obtained from T1w MRI pipeline. Maps were smoothed using Gaussian kernel with 8mm full width at half maximum (FWHM). The resulting maps were remasked by the tissue masks. We also tested the influence of the kernel size from no smoothing to up to 12mm. Thus voxels in GM, WM or GM+WM tissue maps were used as voxel-wise features for diffusion MRI experiments. Regional features correspond to the average value (gray matter density for T1w MRI; FA or MD for diffusion MRI) computed in a set of regions of interest (ROIs) obtained from different atlases. AAL2 atlas containing 120 ROIs (Rolls et al. 2015) was used for T1w MRI. Two JHU atlases, ICBM-DTI-81 white-matter labels atlas (referred as JHULabel with 48 ROIs) and JHU white-matter tractography atlas with a 25% threshold (referred as JHUTract25 with 20 ROIs), were used for diffusion MRI. The different features are shown in Table 4.

**Table 4.** Summary of the different types of features. Both regional and voxel-based feature were extracted with T1w MRI and diffusion MRI.

| Modality | Feature Type | Feature |
| --- | --- | --- |
| Diffusion MRI | Voxel-wise | WM-FA |
| | | WM-MD |
| | | GM-FA |
| | | GM-MD |
| | | WM+GM-FA |
| | | WM+GM-MD |
| | Region-wise | JHULabel-FA/MD |
| | | JHUTract25-FA/MD |
| T1w MRI | Voxel-wise | GM-Density |
| | Region-wise | AAL2 |

## 4.4 Classification

Classification was performed using a linear SVM for both voxel-wise and regional features. As output of the classification, we reported the balanced accuracy, AUC, accuracy, sensitivity, specificity. Additionally, the optimal margin hyperplane (OMH) coefficient maps were reported. The OMH coefficient map represents the influence of each voxel or region on the classification performance. Thus, it characterizes the potential anatomical patterns associated to a given classifier (Cuingnet et al. 2013).

## 4.5 Cross-validation

As emphasized in the recent literature (Varoquaux et al. 2017), it is important to properly perform the cross-validation (CV) procedures. In the present work, the CV procedure included two nested loops: an outer loop evaluating classification performance and an inner loop used to optimize the hyperparameters of the model (C for SVM). More precisely, repeated random splits (all of them stratified) with 250 repetitions was used for outer CV. For hyperparameter optimization, we used an inner loop with 10-fold CV. For each fold, the hyperparameter value corresponding to the highest validation balanced accuracy (BA) is selected, and then these selected hyperparameters are averaged across folds to profit from model averaging. Naturally, the repeated random splits result in different subjects in training and test datasets. For full reproducibility of the current experiments, we provide the tsv and pickle files containing information about the 250 splits of all experiments.

When FR or FS is performed, it is crucial that they are adequately incorporated into the CV procedure. FR is the process of scaling different magnitudes of features to have unit norm and has been widely used in machine learning (Ben-Hur and Weston 2010; Chang and Lin 2011). The features that we used in our study are, in principle, inherently normalized even though not of unit norm, but it still seemed worthwhile to assess the influence of FR. On the other hand, FS aims to identify task-relevant features and thereby reduce the dimensionality to mitigate overfitting (Bermingham et al. 2015). In the present work, we aim to explore the impact of FR and FS on classification performance.

W adopted the commonly used z-score standardization method for FS and incorporated it into the nested CV procedure. Two different FS algorithms were applied: an ANOVA univariate test and an embedding SVM recursive feature elimination (SVM-RFE) (Chandrashekar and Sahin 2014; Guyon et al. 2002). Specifically, the ANOVA test can be seen as a filter without taking the classifier into account and was performed for each feature independently. SVM-RFE uses the coefficients from the SVM models to assess feature importance. Then the least important features, which have the least effect on classification, are iteratively pruned from the current set of features. The remaining features are kept for the next iteration until the desired number of features has been obtained. For both methods, we tested for varying numbers of selected features (1% of the total number of features and then from 10% to 100%, increasing by 10% at each step).

## 4.6 Classification experiments

Four different classification tasks were considered: CN vs AD, CN vs pMCI, sMCI vs pMCI and CN vs MCI.

We assessed the influence of key components on classification performance. First of all, we studied the influence of FR for both DTI and T1w features. Secondly, we compared the performance obtained with different DTI metrics (FA, MD), different feature types (voxel, regional) and different atlases. Thirdly, we studied the impact of imbalanced data. Three tasks (i.e., CN vs pMCI, CN vs MCI and sMCI vs pMCI) have imbalanced data: the number of subjects of the majority group is nearly twice as many as that of the minority group. To assess the impact of data imbalance, a random down-sampling technique was used for each task. In each iteration of the outer CV, this technique randomly excluded certain subjects from the majority group to ensure the subject balance between groups (Raamana 2017). Fourthly, the influence of the degree of smoothing on classification performance was evaluated on the task of CN vs AD. Fifthly, the single contrast and multicontrast registration were compared on the task of CN vs AD. Sixthly, we evaluated the effect of FS bias on the task of CN vs AD. Lastly, we compared the classification performance between diffusion MRI and T1w MRI for task CN vs AD. Of note, the nested CV procedure, in each iteration, guaranteed the same subjects for data split (i.e., training and testing data) between modalities (Raamana 2017).

# 5. Results

Here, we present the classification results with BA as performance metric. All the results with other performance metrics are available at https://github.com/aramis-lab/AD-ML.

## 5.1 Influence of the feature rescaling

Classification results without and with FR are both shown in Table 5. Overall, for DTI features, FR substantially improved the classification results. Specifically, when FR was used and properly incorporated in the CV procedure, the highest BA changed from 0.76 to 0.82 for task CN vs AD, from 0.62 to 0.70 for CN vs pMCI , and from 0.53 to 0.60 for sMCI vs pMCI. Surprisingly, the erroneous classification for region-wise MD features (BA=0.5 with variance=0) were improved by the FR technique (BA>0.7). Due to this positive impact with DTI features, FR was performed in the subsequent experiments. For T1w MRI features, the influence of FR was negligible and resulted in similar results.

**Table 5.** Results of all the classification experiments without and with feature rescaling (FR), respectively. Balanced accuracy (BA) was used as performance metric. Values are presented as mean ± standard deviation (SD).

*For classification without feature rescaling, the mean BA of 0.5 without variation across the 250 splits was caused by the fact that the trained SVM always predict all the cases into the same group. Consequently, this always mutually and exclusively results in sensitivity and specificity being 1 or 0.

| Imaging modality | Feature | FR | CN vs AD | CN vs pMCI | sMCI vs pMCI | CN vs MCI |
|---|---|---|---|---|---|---|
| Diffusion MRI | WM-FA | Without/ With | 0.74±0.089/ 0.75±0.086/ | 0.52±0.109/ 0.57±0.108 | 0.44±0.084/ 0.47±0.074 | 0.57±0.089/ 0.53±0.078 |
| | WM-MD | | 0.71±0.085/ 0.79±0.079 | 0.54±0.088/ 0.66±0.115 | 0.49±0.050/ 0.60±0.105 | 0.59±0.070/ 0.59±0.091 |
| | GM-FA | | 0.70±0.098/ 0.73±0.095 | 0.60±0.112/ 0.61±0.115 | 0.47±0.102/ 0.50±0.082 | 0.58±0.092/ 0.56±0.084 |
| | GM-MD | | 0.76±0.090/<br>0.82±0.082 | 0.62±0.123/<br>0.68±0.112 | 0.52±0.103/<br>0.57±0.102 | 0.60±0.090/<br>0.52±0.084 |
| | WM+GM-FA | | 0.71±0.100/<br>0.73±0.101 | 0.61±0.116 /<br>0.63±0.114 | 0.46±0.091/<br>0.50±0.081 | 0.58±0.089/<br>0.57±0.081 |
| | WM+GM-MD | | 0.76±0.093/<br>0.81±0.081 | 0.61±0.125/<br>0.63±0.115 | 0.53±0.099/<br>0.58±0.109 | 0.59±0.088/<br>0.54±0.086 |
| | JHULabel-FA | | 0.70±0.093/<br>0.76±0.095 | 0.53±0.101/<br>0.53±0.118 | 0.48±0.095/<br>0.48±0.111 | 0.57±0.075/<br>0.54±0.088 |
| | JHULabel-MD | Without/<br>With | 0.50±0*/<br>0.74±0.091 | 0.50±0*/<br>0.54±0.119 | 0.50±0*/<br>0.55±0.108 | 0.50±0*/<br>0.57±0.090 |
| | JHUTract25-FA | | 0.65±0.097/<br>0.61±0.098 | 0.55±0.106/<br>0.52±0.115 | 0.46±0.083/<br>0.51±0.106 | 0.56±0.081/<br>0.54±0.089 |
| | JHUTract25-MD | | 0.50±0*/<br>0.71±0.091 | 0.50±0*/<br>0.70±0.122 | 0.50±0*/<br>0.55±0.119 | 0.50±0*/<br>0.55±0.090 |
| T1w MRI | GM-Density | Without/<br>With | 0.88±0.070/<br>0.88±0.070 | 0.74±0.118/<br>0.74±0.117 | 0.65±0.118/<br>0.65±0.119 | 0.59±0.090/<br>0.58±0.092 |
| | AAL2 | | 0.86±0.078/<br>0.85±0.070 | 0.71±0.126/<br>0.73±0.117 | 0.66±0.118/<br>0.63±0.119 | 0.60±0.086/<br>0.58±0.092 |

## 5.2 Influence of the type of features

Generally, for experiments with FR (Table 5), voxel-wise features provided higher accuracies than regional features. While the difference was moderate for FA, it was more evident for MD. In general, for voxel-wise features, the performance obtained with MD outperformed FA features. For instance, for task CN vs AD, the highest BA was 0.82 for GM-MD features and 0.75 for WM-FA features. Finally, for MD, the inclusion of GM (either in isolation or when combined with WM) considerably increased the performance over the use of WM alone.

## 5.3 Influence of the imbalanced data

For voxel-wise classification, compared to the results using imbalanced data, balanced data showed comparable accuracies for all three tasks, as shown in Figure 2. For the region-wise approach, switching from imbalanced data to balanced data, BA considerably increased from 0.49 to 0.57 with JHUTract25 MD features for task CN vs MCI, and from 0.50 to 0.57 with JHUTract25 MD features for task sMCI vs pMCI (full details are provided in online supplementary eTable 1).

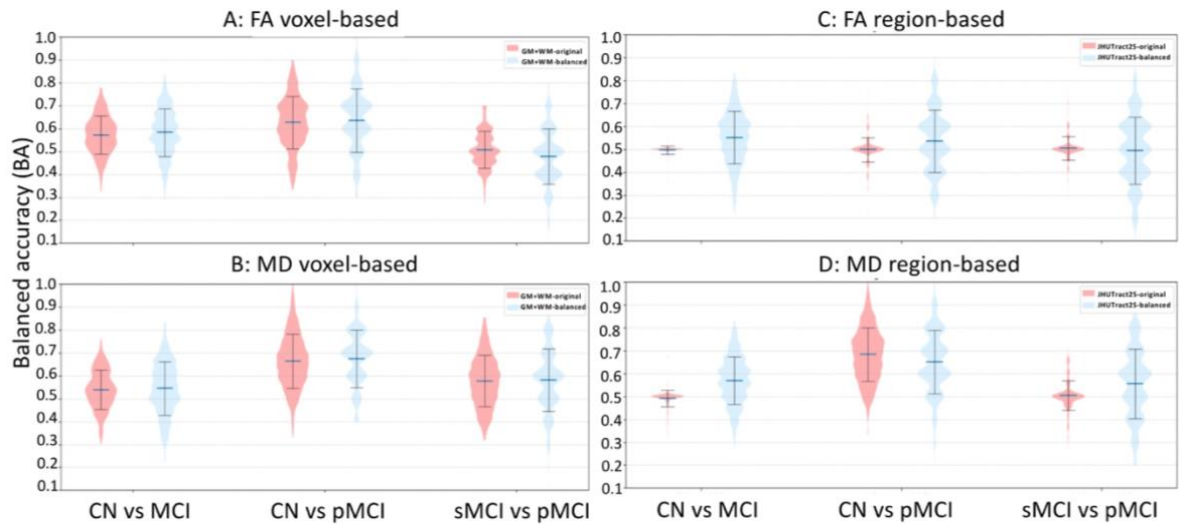

**Figure 2**. Distribution of the balanced accuracy obtained from the randomly balanced classifications (GM+WM-balanced and JHUTract25-balanced) for tasks CN vs MCI, CN vs pMCI and sMCI vs pMCI. For comparison, the original data classification (GM+WM-original and JHUTract25-original) results are also displayed. (A) GM+WM-FA voxel-based feature experiments; (B) GM+WM-MD voxel-based features; (C) JHUTract25 FA regional feature experiments; (D) JHUTract25 MD regional feature experiments.

## 5.4   Influence of the smoothing

We studied the influence of smoothing on classification performance with the following experiments: GM+WM-FA and GM+WM-MD features for task CN vs AD, which are the cases with the highest number of features and for which the performance is higher.

The results are shown in Figure 3. The mean BA increased moderately with the increase of smoothing kernel size for both FA (BA=0.68, 0.72, 0.73 and 0.73) and MD (BA=0.74, 0.80, 0.81, 0.82) features (full details are provided in online supplementary eTable 2).

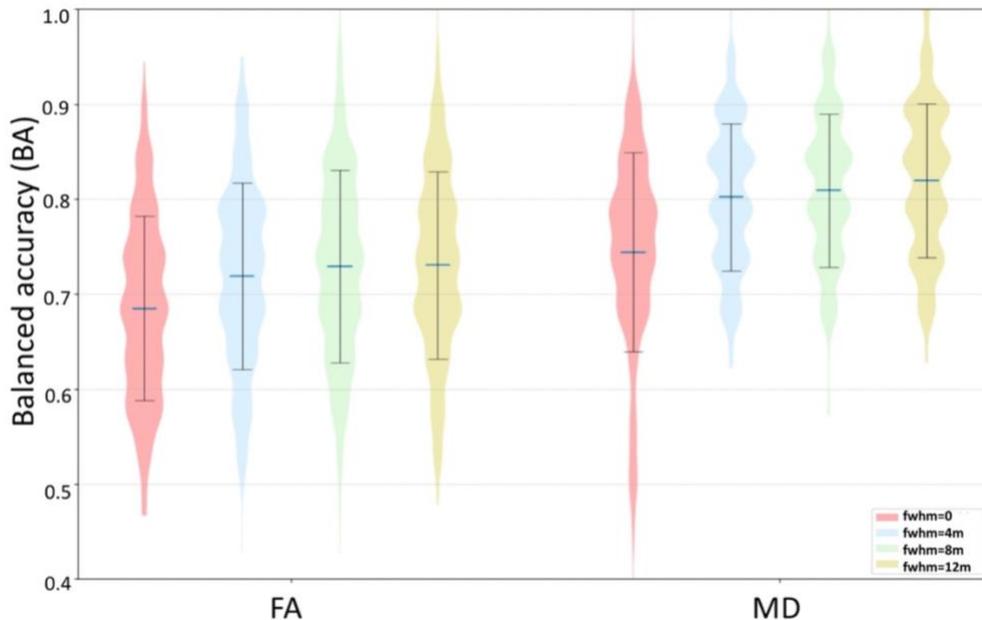

**Figure 3**. Distribution of the balanced accuracy for CN vs AD obtained by varying the degree of smoothing. All experiments were performed with GM+WM features. The influence of the kernel size was tested with different smoothing kernel size, namely no smoothing, 4 mm, 8 mm or 12 mm FWHM.

## 5.5 Influence of the registration method

We assessed the influence of registration (single contrast using FA vs multicontrast using both FA and b0) for voxel-based experiments for task CN vs AD.

As shown in Figure 4, in all cases, the difference in classification performance between single and multicontrast registration was very small (full details are provided in online supplementary eTable 3). When visually checking the registration results, we found that, using single contrast registration, the WM was relatively well-matched while this was not the case for GM and CSF. On the contrary, the entire brain has noticeably better registration accuracy when the FA and b0 images are used simultaneously to drive the registration.

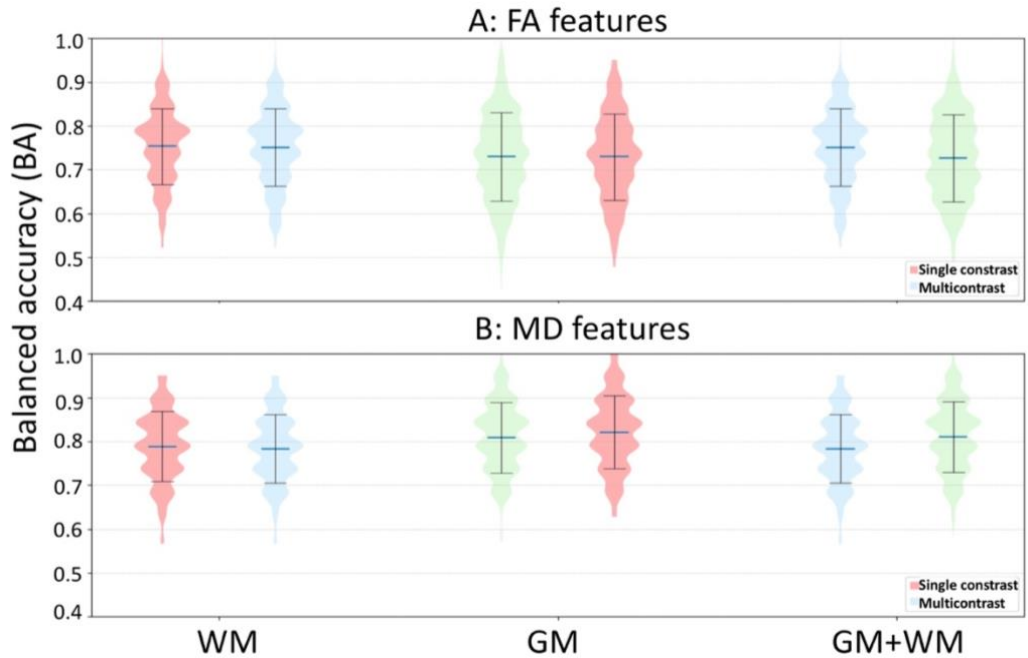

**Figure 4**. Distribution of the balanced accuracy for CN vs AD obtained based on single contrast (FA only) and multicontrast (FA and b0) registration. All experiments were performed with (A) WM-FA, GM-FA and GM+WM feature experiments; (B) WM-MD, GM-MD and GM+WM-MD feature experiments.

## 5.6 Influence of the feature selection bias

To assess the influence of FS bias, the experiments of GM+WM-FA and GM+WM-MD features for task CN vs AD were performed. Note that FR was not considered for both FS approaches, in order to estimate the bias purely caused by non-nested FS.

For both FS algorithms, the non-nested approach resulted in vastly over-optimistic evaluations of performance, from 5% up to 40% increase in BA. Specifically, for ANOVA, the highest BA was obtained with the first 1% most informative voxels for the non-nested approach (BA=0.77 for FA and BA=0.82 for MD), and with all available voxels for the nested approach (BA=0.71 for FA and BA=0.76 for MD). For SVM-RFE, the highest BA was achieved with the first 10% most informative voxels for the non-nested approach (BA=1 for FA and BA=0.82 for MD), and with the first 90% most informative voxels with FA (BA=0.75) and the first 1% most informative voxels with MD (BA=0.77) for the nested approach. Compared to the non-FS case, where all voxels were considered, the nested SVM-RFE FS, not ANOVA, showed potential to improve classification performance (Figure 5).

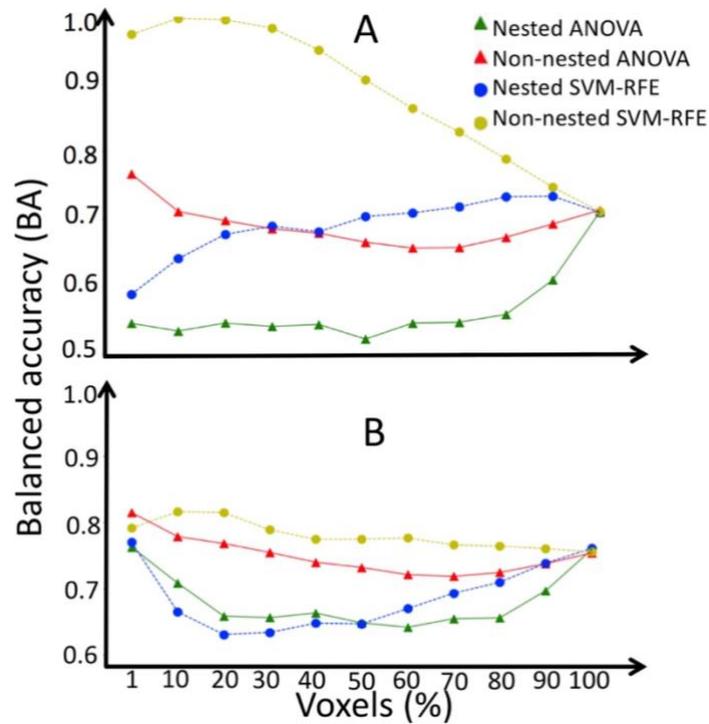

**Figure 5**. Balanced accuracy for CN vs AD obtained by varying the number of voxels for ANOVA and SVM-RFE FS methods. For each FS method, both nested and non-nested method was tested. (A) GM+WM-FA feature experiments; (B) GM+WM-MD feature experiments.

## 5.7 Influence of the imaging modality

Diffusion MRI voxel-wise features and T1w MRI features (i.e., GM-density features) were compared with consideration of both FR and FS (i.e., SVM-RFE) for task CN vs AD, due to the fact that both have positive impact on classification performance.

Overall, the highest BA was obtained with the first 10% most discriminative voxels for GM-density features (BA=0.94) and for GM+WM-MD (BA=0.94). Inside GM voxels, GM-MD features obtained lower performance (BA=0.90) with the first 30% most discriminative voxels than GM-density features (BA=0.94). When no FS was performed, GM-density (BA=0.88) outperformed DTI voxel-wise features (BA=0.82 for GM-MD features) (Figure 6).

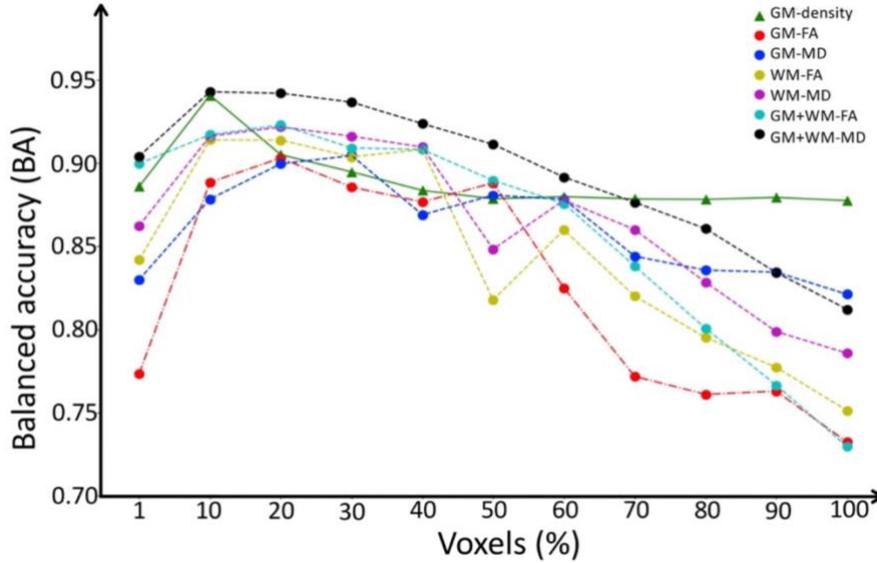

**Figure 6.** Distribution of the balanced accuracy obtained from both T1w and diffusion MRI for tasks CN vs AD. GM density features and DTI voxel-wise features were used for T1w MRI and diffusion MRI, respectively.

## 5.8 Comparison to state-of-the-art results

We aim to present the comparison between our results and the state-of-the-art results. For the current study, we presented results with two different scenarios: i) classification without FR and FS and ii) classification with both FR and FS. For the purpose of a fair comparison, we presented the studies from the literature with following criteria: i) only the studies without data leakage from Table 1 were included, ii) the DTI metric results were reported in the case of the availability of multimodal approaches, and iii) if possible, the balanced accuracy was calculated and presented for each study. Otherwise, accuracy was reported.

Table 6 shows the results of the present study side-by-side with those of the state of the art. The results without FR and FS are in general comparable with those studies using ADNI data, but substantially lower than that of several studies using other datasets (for example EDSD, DZNE, MCXWH, TJH). On the other hand, the results with the adoption of FR and FS obtain better results than those reported in literature (Table 6).

**Table 6**. Comparison of our results to the state-of-the-art using DTI-based features.
ACC: accuracy; BA: balanced accuracy; --: not applicable.

Wen 31

* The results obtained without FR and FS (Table 5).

** The results obtained with both FR and FS. The highest BA was chosen during FS evaluation with different threshold for the number of chosen voxels. For task CN vs AD, the best performance (BA=0.94) was obtained with the first 10% most discriminative voxels for GM+WM-MD features; for task CN vs MCI, the best performance (BA=0.74) was obtained with the first 20% most discriminative voxels for GM+WM-FA features; for task sMCI vs pMCI, the best performance (BA=0.80) was obtained with the first 1% most discriminative voxels for GM+WM-MD features.

| Study | Database | Performances | | |
|---|---|---|---|---|
| | | CN vs AD | CN vs MCI | sMCI vs pMCI |
| Current study* | ADNI | BA=0.76 | BA=0.59 | BA=0.53 |
| Current study** | | BA=0.94 | BA=0.74 | BA=0.80 |
| (Ahmed et al. 2017) | ADNI | BA=0.79 | BA=0.68 | -- |
| (Schouten et al. 2016) | PRODEM | BA=0.74 | -- | -- |
| (Cui et al. 2012) | SMA | -- | BA=0.65 | -- |
| (Dyrba et al. 2013) | EDSD | BA=0.83 | -- | -- |
| (Lella et al. 2017) | ADNI | BA=0.78 | -- | -- |
| (M. Li et al. 2014) | TJH | BA=0.88 | -- | -- |
| (Y. Xie et al. 2015) | MCXWH | -- | BA=0.79 | -- |
| (Zhang and Liu 2018) | ADNI | ACC=0.88 | -- | -- |
| (Maggipinto et al. 2017) | ADNI | ACC=0.76 | ACC=0.60 | -- |
| (Dyrba, Barkhof, et al. 2015) | EDSD | -- | BA=0.76 | BA=0.67 |
| (Dyrba, Grothe, et al. 2015) | DZNE | BA=0.85 | -- | -- |

## 5.9 Potential anatomical pattern

Figure 7 displays the OMH coefficient maps for the most successful task CN vs AD without considering FR or FS. For MD features, discriminative voxels were mainly within the gray matter (hippocampus and medial temporal cortex) (Figure 7B). When restricting the analysis to WM, only small regions were discriminative and outside of the JHUTract25 atlas (Figure 7D). This is consistent with the poor performance obtained with MD regional features. For GM-density features (Figure 7C), the discriminative voxels also included these regions but were more extended (including some regions in the lateral temporal cortex and in the parietal and frontal lobes). For FA, discriminative voxels included both GM and WM regions (Figure 7A). In the GM, discriminative voxels were mainly located within the medial temporal lobe. In the WM, they were more diffuse and absent in deep WM. These regions were close to the forceps minor and major tracts and inferior fronto-occipital fasciculus.

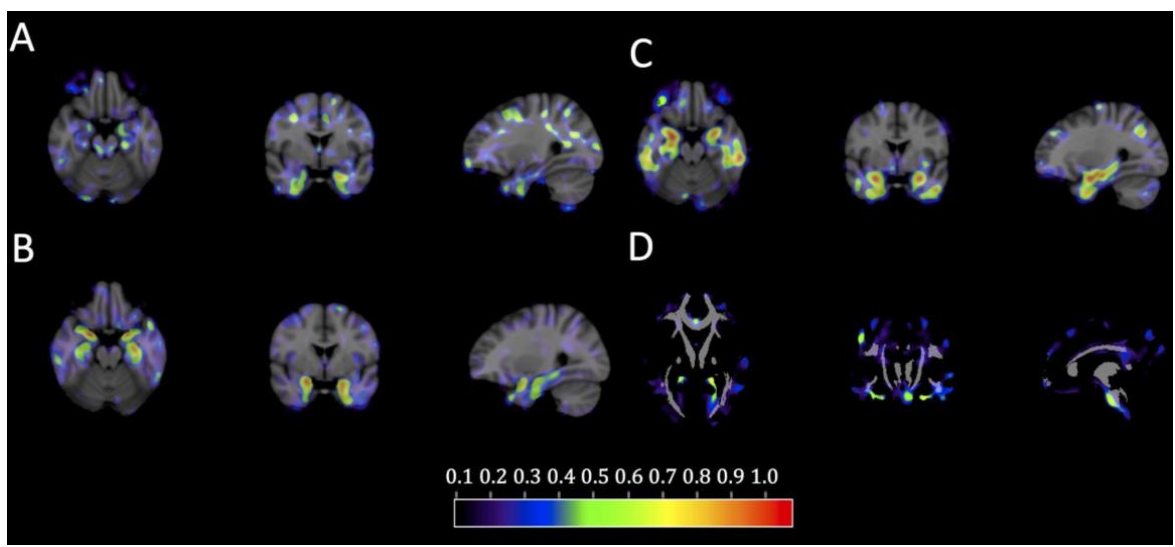

**Figure 7.** Normalized coefficient maps in MNI space. Task CN vs AD with (A) GM+WM-FA features; (B) GM+WM-MD features; (C) GM-Density features; (D) WM-MD features superimposed onto the JHUTract25 atlas (in gray). Warm colors, it means higher likelihood of classification into AD.

# 6. Discussion

In the present work, we proposed an open-source framework for the reproducible evaluation of AD classification from diffusion MRI. It extends our previous framework devoted to T1w MRI and PET by adding conversion of diffusion MRI to BIDS and providing standardized image processing pipelines for diffusion MRI. We then demonstrated its use to assess the influence of key components on classification performance: i) FR, ii) feature types, iii) data imbalance, iv) smoothing strength, v) registration methods (single contrast and multicontrast), vi) FS and vii) imaging modalities (T1w MRI and diffusion MRI).

Generally, we hopefully contribute to make evaluation of machine learning approaches in AD more reproducible and more objective. Firstly, providing the tools to fully automatically convert original ADNI diffusion MRI into the community standard BIDS, we hope to facilitate the future work of researchers. Secondly, the literature (Cuingnet et al. 2011; Lu and Weng 2007; Uchida 2013) suggested that image processing procedures, including steps such as preprocessing, parcellation, registration and intensity normalization, have a strong influence on classification results. Thus, a standard diffusion MRI processing pipeline was proposed in the present work. Lastly, we proposed rigorous CV procedures following recent best practices (Varoquaux et al. 2017). The key components are publicly available in Clinica, a freely available software platform for clinical neuroscience research studies. We hope this framework will allow researchers to easily and rigorously evaluate their own classification algorithms, FS algorithms or image processing pipelines. Furthermore, the modular nature of the framework allows using its different parts in independent ways. For instance, extracted features could be used as input to other ML tools for neuroimaging such, as Neuropredict (Raamana 2017).

We then aimed to benchmark classification performance for future studies. When FR and FS were not performed, the results obtained in our framework were in line with the state-of-the-art results (Table 6*). In our experiments, we obtained the balanced accuracy of 0.76 for task CN vs AD, 0.60 for task CN vs MCI and 0.61 for task CN vs pMCI. One can note that several studies using DTI-based features reported superior performance over our work (Dyrba et al. 2013; Dyrba, Grothe, et al. 2015; M. Li et al. 2014). This was in particular the case for studies using other datasets (such as for instance EDSD, DZNE, MCXWH, TJH). This discrepancy may come from the differences in image quality in different datasets or data sample size. Interestingly, when combining FR and FS, we obtained better classification results than the literature (Table 6**). Specifically, we obtained the BA of 0.94 for task CN vs AD, 0.74 for task CN vs MCI and 0.80 for task sMCI vs pMCI.

We first studied the impact of FR. FR showed a strong positive impact on classification performance for diffusion MRI features, which was not the case for T1w MRI features. Specifically, compared to the case where no FR was performed, FR increased the classification results from 0.76 to 0.82 in BA for task CN vs AD, from 0.62 to 0.70 for task CN vs pMCI, from 0.53 to 0.60 for task sMCI vs pMCI (Table 5). In fact, the features considered in our study have an inherent normalization. So it was not clear that FR would have such an impact on performance. The classification improvement may be due to the fact that FR influences the Euclidean distance between support vectors and, thus, affect the final decision boundary. Although SVMs are known to be sensitive to the way features are scaled (Ben-Hur and Weston 2010; Chang and Lin 2011) and some previous studies have studied its impact. Chapelle and authors (Chapelle et al. 2002) demonstrated the importance of FR in SVM from the statistical learning point of view. Zhang and Wu (Wu and Zhang 2011) empirically evaluated different FR methods' influence on task performance. However, there are scarcities and ambiguities regarding explicit exposition of FR in the literature of AD classification, thus making it hard to objectively evaluate and reproduce their results.

Different types of DTI-based features were assessed (Table 5 and Figure 6). Generally, voxel-wise features provided higher accuracies than region-wise features. This was consistent with a previous study (Demirhan et al. 2015), which reported accuracies of 0.75 for region-wise classification and of 0.88 for voxel-wise classification. Note that the most discriminative voxels (original features without performing FR or FS) for WM-MD classification are outside the regions of the JHUTract25 atlas. This finding explains the poor performance obtained using MD regional features. Thus, the atlas used for region-wise approaches should be chosen with care. Moreover, our results showed that MD gave higher performance than FA for voxel-wise classification. This finding was in conflict with some of the literature: three previous studies (Dyrba et al. 2013; Lella et al. 2017; Maggipinto et al. 2017) claimed that MD and FA obtained comparable results. One possible explanation for this discrepancy may be due to the fact that those studies did not explicitly perform FR. Their findings were consistent with our results without FR and FS (Table 5).

We evaluated the impact of data imbalance on classification performance. It is commonly agreed that imbalanced data may adversely impact classification performance as the learned model will be biased towards the majority class to minimize the overall error rate (Dubey et al. 2014; Estabrooks 2000; Japkowicz and Others 2000). Efforts have been made to deal with imbalanced data, which could be generally classified as algorithmic level (Akbani et al. 2004) and data level (Dubey et al. 2014). In the current study, for voxel-wise classification, we found that the low accuracies obtained in discriminating pMCI from sMCI or CN are potentially caused by the small sample size, rather than by the imbalanced data. Interestingly, Dubey et al showed that the balanced data obtained by several data resampling techniques gave better results than the imbalanced data using T1w MRI from ADNI (Dubey et al. 2014). Thus our hypothesis for the limited sample size needs to be further confirmed as more subjects are becoming available.

Spatial smoothing is widely used in MRI data analysis as a preprocessing step. In the case of diffusion MRI, we did not find a major influence of smoothing kernel size. This is consistent with our previous study (Samper-González et al. 2018) on T1w MRI and PET data. However, smoothing does slightly improve classification performance compared to no-smoothing case: 5% up to 7% relative increase in BA (Figure 3). Spatial smoothing is beneficial for suppressing spatial noise, which is the case of diffusion MRI. Moreover, although different in application, previous studies demonstrated the positive effect of spatial smoothing on quantitative analysis (Chen and Calhoun 2018). Thus we suggest a 4-12 mm FWHM spatial smoothing in the context of classification of AD.

We studied the effect of registration methods as the accuracy of image registration significantly influences the quantitative analysis results. Our experiments showed that multicontrast registration (FA and b0) was substantially more accurate than single contrast (FA only) registration, in particular outside of the white matter, consistently with the conclusions made by (Ceritoglu et al. 2009). This could be expected because FA does not provide much information outside the WM. However, the substantial difference in registration quality did not lead a major difference in classification performance. This may be due to the fact that most of the discriminative voxels are in the vicinity of the WM or in areas where inter-individual variability is lower, such as the hippocampus.

In the literature, researchers have emphasized that "double-dipping" or data leakage or data leakage, referring to the use of test subjects in any part of the training process, such as non-nested FS in the context, is bad practice and may lead to over-fitted classification (Kriegeskorte et al. 2009; Rathore et al. 2017). Similarly, in a recent study, Maggipinto et al showed that the adoption of FS strategies should be taken with care (Maggipinto et al. 2017). They proved that a biased FS, usually a non-nested FS, leads to over-optimistic results. Unfortunately, many previous studies using diffusion MRI for AD classification adopted the non-nested FS and reported nearly perfect classification (Demirhan et al. 2015; Graña et al. 2011; Mesrob et al. 2012; O'Dwyer et al. 2012). In our literature search, we found that 15 out of 34 papers have the potential of data leakage caused by the FS bias (Table 1 and 2). In the current study, our findings reinforced the message that non-nested FS could result in over-optimistic results. With the adoption of the non-nested SVM-RFE FS, nearly perfect performance was achieved. Besides, FA outperformed MD for classification accuracies for this non-nested FS approach. Similar patterns were also witnessed in the study of Maggipinto et al (Maggipinto et al. 2017). Replacing the non-nested FS with the nested one, we obtained considerably inferior performance. On the other hand, we found that, with SVM-RFE not with ANOVA, the nested FS could potentially (slightly) improve the performance compared to the case no FS was performed. The difference between ANOVA and SVM-RFE may stem from the fact that ANOVA is performed for each feature (voxel) independently while GM and WM in contiguous voxels are highly correlated (Mechelli et al. 2005). Interestingly, another study found that, with the adoption of ReliefF algorithm, FS improved the classification accuracy up to 8% compared to the non-FS for task CN vs AD (Demirhan et al. 2015). However, they did not give enough details concerning their validation scheme. In particular, it is not clear if they used a nested FS (Demirhan et al. 2015).

We also systematically compared classification performance between T1w MRI and diffusion MRI. Our results showed that T1w MRI outperformed diffusion MRI without FR and FS (Table 5), but DTI features obtained comparable results compared to T1 after adopting FR and FS methods (Figure 6). Several previous studies have compared the performance of these two modalities without explicitly stating the inclusion of FR. Mesrob et al found that T1w MRI outperformed diffusion MRI (accuracy of 0.77 for T1w MRI vs 0.69 for FA from DTI) for task CN vs AD (Mesrob et al. 2012). However, their results were biased due to the adoption of a non-nested FS. Cui et al found superior performance of T1w MRI over diffusion MRI (accuracy of 0.61 for T1w MRI vs 0.54 for FA from DTI) when classifying CN from MCI for both modalities (Cui et al. 2012). Using a predefined hippocampus ROI approach, Ahmed et al obtained comparable accuracies for both modalities for tasks CN vs AD (accuracy of 0.71 for T1w MRI vs 0.72 for MD from DTI) and CN vs MCI (accuracy of O.65 for T1w MRI vs 0.68 for MD from DTI) (Ahmed et al. 2017). Several factors could explain the better performance of T1w MRI when FR were not performed. First, it is controversial but possible that WM degeneration is a secondary degenerative process compared to brain atrophy (Agosta et al. 2011; Xie et al. 2006). Another possibility is that ADNI diffusion MRI acquisitions used within our study do not make use of state-of-the-art methods that impact on image quality. In particular, no fieldmap data is acquired which leads to suboptimal correction of magnetic susceptibility artifacts (Wu et al. 2008). On the other hand, it seems that, when combining FR and FS, DTI features achieve a performance which is comparable to that of T1w MRI. Nevertheless, since anatomical MR images are easier to acquire, we believe that there is currently no clear reason to prefer diffusion MRI over T1 MRI for computer-aided diagnosis of AD.

Visualization of optimal margin hyperplane coefficient maps revealed which voxels contribute the most to discrimination. FA, MD and GM-Density features shared a typical AD anatomical pattern: voxels in hippocampus and temporal lobe showed more discriminative ability in the classification. These findings were consistent with the literature. DTI-based group comparison analyses demonstrated altered FA or MD in the hippocampus (Fellgiebel et al. 2006; Hanyu et al. 1998; Kantarci et al. 2001; Müller et al. 2005, 2007) and in the temporal lobe (Fellgiebel et al. 2005; Hanyu et al. 1998; Head et al. 2005; Stahl et al. 2007). Moreover, the OMH coefficient map displayed a diffuse pattern for WM voxels in our work. Similar patterns of WM voxels were also witnessed in the FS procedure using diffusion MRI (Demirhan et al. 2015; Dyrba et al. 2013).

In the current study, we only focused on SVM as the classifier. In our previous study (Samper-González et al. 2018), we have compared the performance of different classifiers and found that random forests resulted in inferior performances compared to SVM and logistic regression with L2 regularization. Besides, other studies (Aguilar et al. 2013; Cabral et al. 2015; Sabuncu et al. 2014) have found that the features add more influence on the performance than the classifier itself. Nevertheless, studying the impact of different classifiers could be easily performed with the current framework.

The current framework is for general use. Indeed, we applied this framework to ADNI dataset and offered the automatic tools to convert raw ADNI data into BIDS. However, researchers can easily adapt our framework to other datasets, diseases or tasks. In our previous work (Samper-González et al. 2018), we also applied the framework to other public datasets, namely OASIS and AIBL. In the current work, we did not use OASIS and AIBL because they do not contain DWI data. Originally, our framework was developed for classification tasks. It means that researchers can directly use our framework for other classification problems, for instance, classification between different diseases. Moreover, although dedicated to the classification task, our framework can also be used for a regression task. The modular architecture of the current framework facilitates this adaptation. One would only need to replace the classifiers with regression tools. The other parts of the framework (e.g., image processing, feature extraction or CV) could be directly usable.

Our study has the following limitations. First, ADNI diffusion MRI data was not acquired using the state-of-the-art methods which leads to suboptimal image quality. Related works have proven the negative impact of low image quality on MRI analyses (Alexander-Bloch et al. 2016; Reuter et al. 2015; Yendiki et al. 2014). It is thus possible that diffusion MRI acquired using more recent protocols would provide higher classification accuracies. Moreover, our experiments were performed with a limited data sample size. The limitation came from the data currently available in ADNI. In a previous study (Samper-González et al. 2018), we have demonstrated that increased training set size led to increased classification performance. Thus, both limitations can result in inferior classification performance. Furthermore, our study only explored DTI-based features. With a proper CV and FS, more sophisticated features, such as brain tractography- or network-based features, could also be studied. Lastly, unit and functional tests are important to ensure the robustness of the current framework. Such tests have been added to the Clinica software since version 0.3. We plan to expand their scope to the current framework in the future.

# Information sharing statement

Data used in preparation of this article were obtained from the Alzheimer's Disease Neuroimaging Initiative (ADNI) database (adni.loni.usc.edu). ADNI data was automatically converted into the standard Brain Imaging Data Structure (BIDS) format (https://bids.neuroimaging.io/). The code for participant selection, ADNI converter and image preprocessing is distributed within Clinica (http://www.clinica.run/), an open-source platform for reproducible neuroimaging studies. The paper-specific code for running the experiments is gathered at: https://github.com/aramis-lab/AD-ML.

# Acknowledgments

The research leading to these results has received funding from the program "Investissements d'avenir" ANR-10-IAIHU-06 (Agence Nationale de la Recherche-10-IA Institut Hospitalo-Universitaire-6) ANR-11-IDEX-004 (Agence Nationale de la Recherche-11-Initiative d'Excellence-004, project LearnPETMR number SU-16-R-EMR-16), from the European Union H2020 program (project EuroPOND, grant number 666992, project HBP SGA1 grant number 720270), from the joint NSF/NIH/ANR program "Collaborative Research in Computational Neuroscience" (project HIPLAY7, grant number ANR-16-NEUC-0001-01), from Agence Nationale de la Recherche (project PREVDEMALS, grant number ANR-14-CE15-0016-07), from the European Research Council (to Dr Durrleman project LEASP, grant number 678304), from the Abeona Foundation (project Brain@Scale), and from the French government under management of Agence Nationale de la Recherche as part of the "Investissements d'avenir" program, reference ANR-19-P3IA-0001 (PRAIRIE 3IA Institute). J.W. receives financial support from China Scholarship Council (CSC). O.C. is supported by a "Contrat d'Interface Local" from Assistance Publique-Hôpitaux de Paris (AP-HP). N.B. receives funding from the People Programme (Marie Curie Actions) of the European Union's Seventh Framework Programme (FP7/2007-2013) under REA grant agreement no. PCOFUND-GA-2013-609102, through the PRESTIGE programme coordinated by Campus France.

Data collection and sharing for this project was funded by the Alzheimer's Disease Neuroimaging Initiative (ADNI) (National Institutes of Health Grant U01 AG024904) and DOD ADNI (Department of Defense award number W81XWH-12-2-0012). ADNI is funded by the National Institute on Aging, the National Institute of Biomedical Imaging and Bioengineering, and through generous contributions from the following: AbbVie, Alzheimer's Association; Alzheimer's Drug Discovery Foundation; Araclon Biotech; BioClinica, Inc.;

Biogen; Bristol-Myers Squibb Company; CereSpir, Inc.; Cogstate; Eisai Inc.; Elan Pharmaceuticals, Inc.; Eli Lilly and Company; EuroImmun; F. Hoffmann-La Roche Ltd and its affiliated company Genentech, Inc.; Fujirebio; GE Healthcare; IXICO Ltd.; Janssen Alzheimer Immunotherapy Research & Development, LLC.; Johnson & Johnson Pharmaceutical Research & Development LLC.; Lumosity; Lundbeck; Merck & Co., Inc.; Meso Scale Diagnostics, LLC.; NeuroRx Research; Neurotrack Technologies; Novartis Pharmaceuticals Corporation; Pfizer Inc.; Piramal Imaging; Servier; Takeda Pharmaceutical Company; and Transition Therapeutics. The Canadian Institutes of Health Research is providing funds to support ADNI clinical sites in Canada. Private sector contributions are facilitated by the Foundation for the National Institutes of Health (www.fnih.org). The grantee organization is the Northern California Institute for Research and Education, and the study is coordinated by the Alzheimer's Therapeutic Research Institute at the University of Southern California. ADNI data are disseminated by the Laboratory for Neuro Imaging at the University of Southern California.

# Reproducible evaluation of diffusion MRI features for automatic classification of patients with Alzheimer's disease

## Supplementary Material

We present additional methodological explanations and tables in this supplementary material. More specifically, we first describe the dataset used in our study in eMethod 1. We then present how to install and use the current framework in eMethod 2. Lastly, from eTable1 to eTable4, we present the results of the influence of imbalanced data, smoothing and registration method, respectively.

**eMethod 1.** Dataset used in our study
**eMethod 2.** Instructions on how to install the framework and run the experiments
**eTable 1.** Results of all the classification experiments using balanced data
**eTable 2.** Results of all the classification experiments using different degree of smoothing
**eTable 3.** Results of all the classification experiments using different registration methods approaches

**eMethod 1.** Dataset used in our study.

The data used in the preparation of this article were obtained from the Alzheimer's Disease Neuroimaging Initiative database (ADNI) (adni.loni.usc.edu). The ADNI was launched in 2003 as a public-private partnership, led by Principal Investigator Michael W. Weiner, MD. The primary goal of ADNI has been to test whether serial MRI, PET, other biological markers, and clinical and neuropsychological assessment can be combined to measure the progression of MCI and early AD. Over 1,650 participants were recruited across North America during the first three phases of the study (ADNI1, ADNI GO and ADNI2). Around 400 participants were diagnosed with AD, 900 with MCI and 350 were control subjects. Three main criteria were used to classify the subjects (Petersen et al. 2010). The normal subjects had no memory complaints, while the subjects with MCI and AD both had to have complaints. CN and MCI subjects had a mini-mental state examination (MMSE) score between 24 and 30 (inclusive), and AD subjects between 20 and 26 (inclusive). The CN subjects had a clinical dementia rating (CDR) score of 0, the MCI subjects of 0.5 with a mandatory requirement of the memory box score being 0.5 or greater, and the AD subjects of 0.5 or 1. The other criteria can be found in (Petersen et al. 2010).

**eMethod 2.** Instructions on how to install the framework and run the experiments.

We assume that you have installed all the dependencies of the Clinica software platform (http://www.clinica.run) and downloaded the ADNI original data without changing anything. Here the command lines for reproducing our experiments:

Step 1, go back to paper version of Clinica.

*bash 1-go_back_to_paper_version_clinica.sh*

Step 2, convert the original ADNI data into BIDS format. The original ADNI data should be downloaded without further touch (Data we used in our paper was downloaded in October 2016).

*bash 2-ADNI_conversion.sh*

Step 3, create the subjects lists.

*python 3-create_subjects_list.py*

Step 4, create the demographic table information.

*python 4-population_statistics.py*

Step 5, run image processing pipelines. The fist preprocessing and processing pipelines were already integrated into Clinica software and the postprocesing python script was in the current repository. Run these pipelines sequentially:

*bash 5-ADNI_preprocessing.sh*
*bash 6-ADNI_processing.sh*
*python 7-ADNI_postprocessing.py*

Step 6, run classification tasks.
Classification results of original data on T1-weighted and diffusion MRI:

*python 8-ADNI_classification_original_data.py*

Classification results of balanced data on diffusion MRI:

*python 9-ADNI_classification_balanced_data.py*

Classification results of feature selection bias on diffusion MRI:

*python 10-ADNI_classification_feature_selection.py*

**eTable 1.** Results of all the classification experiments using balanced data. Balanced accuracy was used as performance metric.

| Imaging Modality | Feature | CN vs pMCI | sMCI vs pMCI | CN vs MCI |
|---|---|---|---|---|
| Diffusion MRI | WM-FA | 0.58±0.151 | 0.48±0.150 | 0.54±0.113 |
| | WM-MD | 0.70±0.140 | 0.62±0.138 | 0.58±0.090 |
| | GM-FA | 0.63±0.137 | 0.48±0.151 | 0.57±0.1073 |
| | GM-MD | 0.66±0.144 | 0.58±0.146 | 0.54±0.101 |
| | WM+GM-FA | 0.64±0.146 | 0.48±0.156 | 0.58±0.110 |
| | WM+GM-MD | 0.67±0.139 | 0.58±0.150 | 0.55±0.105 |
| | JHULabel-FA | 0.51±0.138 | 0.49±0.138 | 0.54±0.101 |
| | JHULabel-MD | 0.60±0.088 | 0.54±0.142 | 0.59±0.078 |
| | JHUTract25-FA | 0.53±0.135 | 0.49±0.142 | 0.55±0.118 |
| | JHUTract25-MD | 0.65±0.148 | 0.56±0.144 | 0.57±0.103 |

**eTable 2.** Results of all the classification experiments using different degree of smoothing. Balanced accuracy was used as performance metric.

| Imaging Modality | Feature | Smoothing (mm) | CN vs AD |
|---|---|---|---|
| Diffusion MRI | WM+GM-FA | 0 | 0.68±0.095 |
| | | 4 | 0.72±0.098 |
| | | 8 | 0.73±0.099 |
| | | 12 | 0.73±0.098 |
| | WM+GM-MD | 0 | 0.74±0.096 |
| | | 4 | 0.80±0.090 |
| | | 8 | 0.81±0.092 |
| | | 12 | 0.82±0.099 |

**eTable 3.** Results of all the classification experiments using different registration methods. Balanced accuracy was used as performance metric.

| Imaging modality | Registration method | Feature | CN vs AD |
|---|---|---|---|
| Diffusion MRI | Single modal | WM-FA | 0.75±0.086 |
| | | WM-MD | 0.78±0.079 |
| | | GM-FA | 0.73±0.095 |
| | | GM-MD | 0.82±0.082 |
| | | WM+GM-FA | 0.73±0.101 |
| | | WM+GM-MD | 0.81±0.081 |
| | Multimodal | WM-FA | 0.75±0.098 |
| | | WM-MD | 0.78±0.097 |
| | | GM-FA | 0.73±0.094 |
| | | GM-MD | 0.82±0.094 |
| | | WM+GM-FA | 0.73±0.095 |
| | | WM+GM-MD | 0.81±0.099 |